# Ungated, plug-and-play preclinical cardiac CEST-MRI using radial FLASH with segmented saturation

Jonah Weigand-Whittier[1], Michael Wendland[2], Bonnie Lam[1], Mark Velasquez[1], Moriel H. Vandsburger[1,*]

1 Department of Bioengineering, University of California Berkeley, Berkeley, CA, USA
2 Berkeley Preclinical Imaging Core, University of California Berkeley, Berkeley, CA, USA

* Corresponding author:
  **Name** Moriel H. Vandsburger
  **Department** Department of Bioengineering
  **Institute** University of California Berkeley
  **Address** 284 Hearst Memorial Mining Building
  Berkeley, CA 94270
  USA
  **Email** moriel@berkeley.edu

**Abstract**

**Purpose**: ECG and respiratory-gated preclinical cardiac CEST-MRI acquisitions are difficult due to variable saturation recovery with $T_1$, RF interference in the ECG signal, and offset-to-offset variation in Z-magnetization and cardiac phase introduced by changes in cardiac frequency and trigger delays.

**Methods**: The proposed method consists of segmented saturation modules with radial FLASH readouts and golden angle progression. The segmented saturation blocks drive the system to steady-state, and because center k-space is sampled repeatedly, steady-state saturation dominates contrast during gridding and reconstruction. Ten complete Z-spectra were acquired in healthy mice using both ECG and respiratory-gated, and ungated methods. Z-spectra were also acquired at multiple saturation $B_1$ values to optimize for amide and creatine contrasts.

**Results**: There was no significant difference between CEST contrasts (amide, creatine, MT) calculated from images acquired using ECG and respiratory-gated and ungated methods ($p$ = 0.27, 0.11, 0.47). A saturation power of 1.8μT provides optimal contrast amplitudes for both amide and total creatine contrast without significantly complicating CEST contrast quantification due to water direct saturation, magnetization transfer, and RF spillover between amide and creatine pools. Further, variability in CEST contrast measurements was significantly reduced using the ungated radial FLASH acquisition ($p$ = 0.002, 0.006 for amide and creatine respectively).

**Conclusion**: This method enables CEST mapping in the murine myocardium without the need for cardiac or respiratory gating. Quantitative CEST contrasts are consistent with those obtained using gated sequences, and per-contrast variance is significantly reduced. This approach makes preclinical cardiac CEST-MRI easily accessible, even for investigators without prior experience in cardiac imaging.

**INTRODUCTION**

Cardiac magnetic resonance imaging (CMR) is an invaluable tool both clinically and in research and has emerged as a gold standard for the assessment of changes in ventricular structure and function in response to both acute and chronic disease states (1,2). The use of chelated gadolinium and paramagnetic nanoparticle contrast agents has expanded the utility of CMR further, allowing for the assessment of myocardial perfusion and viability (3–5). Recently, changes in cardiac metabolism have been identified as a hallmark and underlying cause of heart failure with preserved ejection fraction(6); reinvigorating interest in the assessment of cardiac metabolism and energetics non-invasively using CMR spectroscopy. However, MR spectroscopic imaging requires long scan times, custom RF hardware, and large voxel sizes due to metabolite concentration relative to water (7,8). Hyperpolarized $^{13}$C MRI remedies the issue of SNR with exogenous $^{13}$C-labeled contrast agents but still requires specialized equipment and additional trained personnel (9). Chemical exchange saturation transfer (CEST) is an emerging molecular imaging method that utilizes endogenous contrast generated by specific proteins, lipids, metabolites, and macromolecules for in vivo imaging of tissue biochemistry (10–12). It is possible to derive several useful MR contrasts from multiple proton exchange pools and indirect exchange mechanisms such as amide (13–16), amine (17,18), relayed nuclear Overhauser enhancement (NOE) (15,19), and magnetization transfer (MT) (20). In the heart, these endogenous contrasts can be used to assess cardiac fibrosis, cardiometabolic disorders, dysregulated cardiac energetics, and treatment response (21–24).

Following explosive growth in neuroimaging with CEST-MRI, Zhou et al. published standards and recommendations for the acquisition and analysis of CEST-MRI of the brain (16). In contrast, cardiac CEST-MRI remains comparatively niche and preclinical cardiac CEST remains particularly challenging. First, the need for robust electrocardiograph (ECG) gating requires skilled preparation. Second, as saturation powers are increased, radiofrequency (RF) interference contaminates the ECG signal, making gating difficult or even impossible (Figure 1). The impact of inconsistent gating and drift in heart rates during imaging induces high variability in the magnitude of $T_1$ relaxation between saturation periods, leading to transient steady-state encoding of CEST contrast in the longitudinal magnetization. Subsequently, the CEST contrast derived from such images will reflect a combination of the underlying macromolecular concentration and the physiological conditions during acquisition. Previous studies have utilized a variety of methods including dummy scans to preserve saturation (25), vendor-specific retrospective self-gating (24), and segmented saturation schemes with electrocardiograph (ECG) and respiratory gating to limit some of the aforementioned problems (26). However, none of these methods can easily be used off the shelf without skilled preparation and detailed knowledge of both CEST and cardiac acquisitions (i.e., not of these sequences could be referred to as "plug-and-play"), and all still result in relatively high variability in quantified CEST contrasts.

As interest in preclinical cardiac CEST-MRI grows, it is critical that a robust and easily implemented acquisition method be established. Further, the resulting data, including quantitative contrast measurements, must be replicable and interpretable across institutions and scanners. In this work, we aim to address these issues by introducing a novel plug-and-play method with an optimized saturation scheme for simultaneous assessment of all relevant endogenous CEST contrasts in the murine myocardium (magnetization transfer, amide, creatine, and NOE) without the need for cardiac or respiratory gating. We also provide open-source image reconstruction and analysis pipelines in Python to promote adoption and facilitate the sharing of results. The results of this study show reduced image acquisition times and accurately reconstructed CEST contrast magnitudes with significantly reduced variability.

**METHODS**

**2.1 Sequence design**

The ungated cardiac CEST-MRI pulse sequence consisted of two repeating modules. First, a short CEST preparation period consisting of 13 frequency-selective Gaussian radiofrequency pulses (duration = 13.7 ms, bandwidth = 200 Hz; interpulse delay = 4.56 ms; dephasing gradient duration = 2 ms) was played out (Figure 2a). After each preparation period, between 4 and 20 $k$-space lines were acquired using a radial FLASH readout scheme (flip angle = 15°, TR = 8.0 ms, TE = 2.104 ms) with a golden angle ($\varphi = 111.5°$) radial progression (27) (Figure 2b,c). These saturation and readout modules were repeated up to 75 times per frequency offset, for up to 3.33x radial oversampling. After each frequency offset image was acquired, an 8 s recovery period was played out to fully reset the longitudinal magnetization. The entire sequence was then repeated for each prescribed frequency offset after an 8 s recovery time, the purpose of which was to allow for the longitudinal magnetization to recover completely between acquisitions.

**2.2 Small animal imaging**

All animal experiments were performed in accordance with the Institutional Animal Care and Use Committee guidelines. Six C67BL6/J mice (two female, four male) were purchased from Jackson Laboratories (Bar Harbor, ME). Mice were anesthetized with 1-3% isoflurane; an internal body temperature of $36.7 \pm 0.55$°C was maintained using a circulating waterbed during all CEST experiments. For acquisitions involving ECG and respiratory gating, physiological monitoring equipment and gating software were used (Small Animal Instruments Inc., Stony Brook, NY). Mice were scanned up to two times each in separate sessions for a total of ten datapoints per experiment type.

In vivo images were acquired using a Bruker PharmaScan 70/16 7T MRI (Bruker, Ettlingen, Germany). To minimize diaphragmatic motion, animals were oriented in the prone position with the cardiac region positioned over a four-channel phased array receive coil (Bruker BioSpin, Billerica, MA) (Figure S1).

A single mid-ventricle short-axis slice (matrix size = 192 × 192, FOV = 25mm × 25mm, slice thickness = 1 mm) was prescribed from a four-chamber long-axis localizer. This slice geometry was inherited by all CEST acquisitions. Complete Z-spectra were acquired both with and without ECG and respiratory gating using a $B_1$ saturation power of 1.1μT at the following frequency offsets: -10 to -5 ppm with a step size of 1 ppm, -4 to 4 ppm with a step size of 0.2 ppm, 5 to 10 ppm with a step size of 0.2 ppm. Reference images were acquired at 200 ppm at the start and end of each acquisition. Reference images were also acquired periodically every seven frequency offset images to account for potential thermal drift. ECG and respiratory-gated images were also acquired using segmented saturation and a radial FLASH readout; however, gated acquisitions were limited to 4 trajectories per readout segment, as worst-case trigger delays had to be accounted for when matching $T_1$ relaxation with ungated acquisitions. During a subsequent imaging session, complete Z-spectra were acquired without ECG and respiratory gating using $B_1$ saturation powers of 1.1, 1.8, 2.4, and 3.1μT.

## 2.3 Data analysis

### 2.3.1 Image reconstruction

All images were reconstructed using custom Python scripts. Images were reconstructed from raw *k*-space and trajectory data by non-uniform fast Fourier transform (28) using the Berkeley Advanced Reconstruction Toolbox (29). Coil images were combined using root-sum-of-squares. Ungated images with Nyquist sampling were retrospectively reconstructed from oversampled images by removing trajectories from the end of each acquisition.

### 2.3.2 CEST contrast analysis

All data were analyzed using custom Python scripts. Regions of interest were calculated from anatomical features based on the American Heart Association's standardized myocardial segmentation guidelines for the left ventricle (LV) mid-cavity short-axis (30). First, epicardial and endocardial bounds were prescribed; then right ventricular insertion points were specified and used to determine myocardial segmentation. The anteroseptal segment was chosen for analysis, as contrast quantitation in inferior and lateral segments suffers due to increased $B_0$ and $B_1$ inhomogeneities despite a favorable receive coil sensitivity profile (21) (Figure S2).

CEST Z-spectra were calculated from average signal intensities over each myocardial segment by normalizing each CEST acquisition to a reference image. To accurately account for frequency drift due to coil element and sample heating, a unique reference image was calculated for each CEST-weighted image by interpolation between intermittently acquired reference images (17).

CEST contrasts were quantified using a five-pool exchange model consisting of water, MT, NOE, amide, and total creatine (creatine and phosphocreatine). Lorentzian line fits were calculated using a two-step least-squares fitting process adapted from Zaiss *et al.* (31). Least-squares fitting parameters are listed in Table 1.

Z-spectra were corrected for $B_0$ inhomogeneity by calculating initial water and MT fits within each myocardial segment and shifting the frequency axis based on the calculated water peak center frequency.

### 2.3.3 Image quality and segmentation analysis

To establish a metric for image quality, frequency offset and reference images were compared in a matrix using the structural similarity index measure (SSIM) (32) such that each image was compared to every other image within each CEST acquisition. The mean value of each matrix, with diagonals excluded, was then calculated to quantify image quality and consistency.

Differences in cardiac phase were calculated using the Dice similarity coefficient (DSC) (33). Within each acquisition, myocardial LV ROIs from the first, last, and middle reference images were compared, and the mean DSC was taken as a measure of similarity in cardiac phase throughout the acquisition.

### 2.3.4 Statistical analysis

All statistical analyses were performed using the open-source SciPy scientific computing library for Python (34). In all box plots, the central horizontal line represents the mean value, the box represents the upper (third) and lower (first) quartiles, and whiskers extend to points within 1.5x the interquartile range. Unfilled points represent outliers. Differences between groups were calculated using the pairwise t-test, differences between variances were calculated using Bartlett's test (35). Differences were considered significant at $p < 0.05$. Statistics in this text are presented as mean ± SD unless otherwise specified.

## RESULTS

### 3.1 Comparison between ECG and respiratory-gated and ungated acquisitions

Representative images and Z-spectra with and without ECG and respiratory gating are shown in Figure 3. Qualitatively, the ungated reference image captures ventricular diastole whereas the gated image appears more systolic. CEST and MT contributions were visibly reduced in the ungated scenario but remain clearly defined in both the gated and ungated Z-spectra.

Box plots representing the distributions of CEST contrasts derived from gated and ungated acquisitions, both with and without poor ECG gating included in the analysis, are shown in Figure 4. Poor ECG gating was defined as >25% missed or extraneous triggers during a predefined observation period at the beginning of each scan.

The mean CEST contrast values for amide, creatine, and MT in the gated scenario were 3.89 ± 2.73%, 7.21 ± 4.43%, and 5.77 ± 3.22% respectively. In the ungated scenario, they decreased slightly to 2.87 ± 0.86%, 4.71 ± 1.61%, and 4.83 ± 2.33% for amide, creatine, and MT respectively. There were no significant differences between CEST contrasts derived from ungated acquisitions and gated acquisitions regardless of whether poorly gated acquisitions were included ($p = 0.27, 0.11, 0.47$ for amide, creatine, and MT respectively) (Figure 4a) or excluded ($p = 0.71, 0.23, 0.33$ for amide, creatine, and MT respectively) (Figure 4b).

Interscan variance decreased per CEST contrast when the ungated method was used. A significant difference in variance between gated and ungated acquisitions was observed for both amide ($p = 0.002$) and creatine ($p = 0.006$) contrasts when acquisitions with poor ECG gating were included in the analysis (Figure 4a). When acquisitions with poor gating were excluded, the differences in variance were no longer significant (Figure 4b). In all cases, the per-contrast variance was increased without radial oversampling.

Image quality and cardiac phase consistency were both significantly improved using the plug-and-play method versus the gated standard (Figure 5). Across all ungated acquisitions, SSIM and DSC were $0.73 \pm 0.03$ and $0.87 \pm 0.03$ respectively, whereas both values fell to $0.57 \pm 0.03$ and $0.78 \pm 0.09$ respectively in the gated scenario. With Nyquist sampling, ungated image quality was significantly worse than both ECG gated and ungated acquisitions with radial oversampling (SSIM = $0.45 \pm 0.04$). Notably, changes in image contrast and homogeneity due to saturation effects make it so that the mean SSIM derived across all images will never approach 1.00; rather, whole image SSIM measured across all CEST-weighted and reference image serves as a proxy for SNR with a small contribution from differences in cardiac and respiratory phase.

**3.2 Optimization of saturation $B_1$ for amide and creatine CEST contrast using ungated acquisition**

The mean CEST contrast values derived from ungated acquisitions at 1.1μT for amide, creatine, and MT were $3.63 \pm 1.46\%$, $6.88 \pm 1.65\%$, and $4.42 \pm 2.05\%$ respectively. When saturation power was increased to 1.8μT contrast values increased to $6.02 \pm 1.55\%$, $7.48 \pm 2.88\%$, and $9.46 \pm 2.66\%$; at 2.4μT, $7.58 \pm 2.12\%$, $8.32 \pm 2.37\%$, and $12.67 \pm 2.03\%$; and finally, at 3.1μT, $7.01 \pm 1.61\%$, $6.69 \pm 1.83\%$, and $19.07 \pm 3.58\%$ for amide, creatine and MT respectively. Creatine contrasts are consistent across all tested saturation powers, whereas both amide and MT increase with saturation power (Figure 6). Representative Z-spectra (Figure 7a,b,c) and Lorentzian difference plots (Figure 7d,e,f) derived from acquisitions at these various saturation powers show clear delineation between amide and creatine peaks at 1.1 and 1.8μT. As saturation powers are increased up to 3.1μT, amide and creatine CEST contrast values decrease, as separation of the Lorentzian line shapes becomes difficult. Representative ungated images capture ventricular diastole with minimal variation; however, spatial inhomogeneities due to direct saturation are more pronounced at higher saturation powers (Figure 8). Further, creatine quantification suffers in inferior and lateral myocardial segments, especially at higher saturation powers (Figure S3).

Representative pixelwise CEST contrast maps for MT, amide, and creatine were derived from ungated images acquired with saturation powers of 1.1 and 1.8μT and compared with contrast maps derived from an ECG gated acquisition in the same animal with a saturation power of 1.1µT (Figure 9). The ungated maps demonstrate greater homogeneity across the anterior and anteroseptal myocardial segments for all CEST contrasts, with decreasing homogeneity across the anterolateral and inferoseptal segments. The inferolateral and inferior segments demonstrate poor contrast homogeneity with significant outliers and several regions in which least-squares Lorentzian fits failed to converge within the given number of iterations. Overall, creatine contrast maps are more homogeneous when using a saturation power of 1.1µT, although amide contrast quantification suffers in all regions.

**DISCUSSION**

In this work, we introduce a novel plug-and-play preclinical acquisition and analysis method for rapid multi-target CEST-CMR without the need for physiological triggering and with reduced variability in quantified contrasts. This acquisition was designed based on the idea that a radial readout scheme would eliminate coherent motion artifacts and that the repeated sampling of the center *k*-space would allow steady-state saturation conditions to dominate image contrast during gridding and reconstruction. In the anesthetized mouse, the mean cardiac phase is diastolic and relatively stationary (i.e., quiescent) without respiratory motion (36). By using readout segments that approximately match the cardiac cycle duration and incorporating a radial oversampling scheme, a mean diastolic image can be reconstructed with minimal respiratory motion artifacts and blurring. This approach eliminates the need for retrospective gating.

The primary challenge when designing this sequence was adjusting the number of readouts per saturation period such that the loss of CEST contrast with $T_1$-relaxation was limited, while scan time was kept within reasonable limits, and blurring due to cardiac motion was reduced as much as possible. We found that if the length of each readout segment was limited to approximately the length of a single cardiac period (~100-150 ms in mice), the majority of *k*-space was acquired during ventricular diastole. Therefore, acquiring approximately 20 readouts per saturation period strikes an ideal balance between CEST contrast, scan length, and image quality.

CEST contrast at 2 ppm downfield from water is broadly attributed to guanidine amine protons, with contributions from both total creatine and protein arginine in vivo (37,38). A study by Haris *et al.* implied that in the myocardium, creatine is the primary contributor to CEST contrast at 2 ppm (22). However, a more recent study by Ayala *et al.* suggests that when using pulsed contrast generation with Gaussian RF pulses at body temperature the contrast at 2.2 ppm is from total creatine (phosphocreatine and creatine) (39) as was also seen at higher fields in skeletal muscle (40). Creatine and amide CEST contrasts obtained using the ungated method were broadly consistent with previous studies. Pumphrey *et al.* derived a mean creatine CEST contrast of approximately 8% using a cartesian readout, ECG and respiratory gating, and an average $B_1$ saturation power of 4.9μT (25). These contrasts were calculated using the magnetization transfer ratio asymmetry (41) as opposed to Lorentzian line fitting. Lam *et al.* derived creatine and amide contrasts of 7.34 ± 1.71% and 4.6 ± 2.49%, respectively, using a radial readout with a similar segmented saturation scheme, ECG and respiratory gating, and a peak $B_1$ of 1.0μT (26). Ultimately, CEST contrasts are only semi-quantitative and it is most important that measurements are consistent and reproducible. As we have shown, quantification can be difficult in studies with prospective ECG gating due to intrascan Z-spectral inconsistencies introduced by changes in the captured cardiac phase and macro-level inconsistencies in longitudinal magnetization resulting from variable ECG and respiratory trigger delays (Figure S4).

Some of the observed differences in the magnitude of CEST contrasts can be attributed to variations in per-segment readout times depending on the cardiac rate. Although we endeavored to match ungated readout times to worst-case gated readout times with maximally long ECG and respiratory trigger delays, the majority of gated readouts are shorter than 20 segment, ungated readouts. Consequently, mean CEST contrast values are lower in the ungated acquisition due to saturation recovery with $T_1$. Although RF interference at higher saturation powers limited our comparison between ECG gated and ungated acquisitions to a saturation power of 1.1μT, we expect that similar trends would be observed regardless of saturation power. Our results suggest that preclinical cardiac CEST experiments utilizing this plug-and-play method should use a saturation power of 1.8μT when performed at 7T. This achieves effective labeling (42) of both amide and creatine based on established exchange rates (13,43,44), without causing significant line broadening that complicates peak separation and Lorentzian fitting. For single-target creatine imaging, a saturation power of 1.1μT provides more homogeneous creatine contrast at the expense of amide contrast amplitude and homogeneity.

The nominal duration for each full ECG gated acquisition was 29.72 min—practically; these acquisitions take longer, 32.22 ± 1.19 min, as trigger delays, are not factored into scan time estimations. With 3.33x radial oversampling and 20 trajectories acquired per readout segment, the duration for each full ungated acquisition was consistently 29.18 min. Ungated acquisitions could potentially be accelerated further with Nyquist sampling (9.73 min) or 2x radial oversampling (19.45 min); however, a penalty to SNR and, therefore, accurate CEST contrast quantification is incurred. The number of trajectories measured per segment could also potentially be optimized on a per-animal basis, although we found that 20 trajectories per segment delivered consistent measurements and image quality across animals and acquisitions. Importantly, even at a level of 3.33x oversampling this plug-and-play method provides a complete Z-spectrum with high analytical quality for simultaneous quantitation of all endogenous CEST contrasts. In contrast, the acquisition of enough frequency offsets required to calculate CEST contrasts for all endogenous targets based on outdated MTR asymmetry measurements using established cartesian ECG and respiratory gating would yield largely similar scan times.

Although radial oversampling is not strictly necessary for CEST contrast quantification, some level of oversampling provides considerable benefits and should be utilized provided scan time is not a limiting factor. Higher oversampling levels improve SNR, simply because more *k*-space data is acquired. *K*-space coverage at higher spatial frequencies is also improved. Moreover, increased radial oversampling leads to a greater proportion of *k*-space being acquired after the system has been driven to steady-state and in periods of ventricular diastole. Although ungated images suffer some level of blurring due to cardiac and respiratory motion, particularly at higher spatial frequencies, radial oversampling mitigates blurring in the myocardium and CEST contrast quantification seems to be unaffected.

It is well established that genetically engineered mouses models of human diseases are relevant for understanding the roles of genes, physiology, and treatment (45,46); however, there is limited data on the use of CEST-MRI in these settings. There are likely significant changes in CEST contrasts associated with dysregulated cardiac energetics and protein production, which could be helpful in developing new understandings of disease pathology and treatment in animal models of hypertrophic cardiomyopathy (47,48), cardiac amyloidosis (49), and heart failure with preserved ejection fraction (50).

In silico experiments are ubiquitous in the field and are often an important step for optimizing CEST acquisition parameters such as saturation time, saturation power, and pulse shape (39,51–53). Due to the relative complexity of the repeated saturation and readout segments employed in this study, simulating spin and exchange dynamics is difficult. If the difficulty of simulating the dynamics could be overcome, it is possible that a similar pulse sequence could be employed for cardiac CEST magnetic resonance fingerprinting (51,54).

Notably, this method also allows for the extraction of physiological signals from *k*-space data (Figure S5), which has been widely employed for retrospective cardiac and respiratory self-gating (55,56) and opens the door to potential simultaneous CEST and cardiac function experiments. Physiological signals were extracted from two z-score normalized center *k*-space points from each readout. High frequency components were then filtered out based on expect cardiac and respiratory rates. Although per-coil respiratory and cardiac signals were selected manually in this case, frequency spectrum analysis could be used to automatically select optimal coils, as described by Zhou *et al*. (55). ECG gated cardiac CEST measurements are significantly easier to perform in the clinical setting using centric-ordered cartesian readouts (21,39); however, a method incorporating segmented saturation and radial readouts could allow for free-breathing, ungated CEST imaging with retrospective reconstruction in pediatric patients or patients with limited ability to perform long breath-holds.

The regional variations in direct saturation (Figure 8) and CEST contrast quantification (Figure S6) are consistent with the spatial inaccuracies due to $B_1$ inhomogeneity across the myocardium identified by AlGhuraibawi *et al.* at 3T (57). Recently, spatial-spectral saturation pulses have been shown to remedy issues with $B_1$ inhomogeneity in cardiac CEST acquisitions, and also exhibit exchange-rate selective properties, improving quantification across the myocardium and allowing individual targeting of creatine and phosphocreatine (39). Future work should explore the feasibility of implementing tailored spectral-spatial pulses in preclinical imaging routines. Importantly, the amplitudes of CEST contrasts depend not only on physiological parameters such as metabolite concentration, chemical exchange rates, and native relaxation times, but also on saturation and acquisition parameters such as mean or peak saturation power, pulse duration, pulse shape, interpulse delay, and number of pulses (58,59). Therefore, for CEST contrasts to be relevant across centers and studies, or across time points in longitudinal studies, it is necessary to develop standardized acquisitions (53). It is also crucial that experiments are reproducible, and we invite the CEST and CMR communities to use the described pulse sequence, as well as the publicly available reconstruction and CEST analysis scripts.

**CONCLUSION**

This work introduces a plug-and-play method for ungated CEST imaging of the murine myocardium, making complicated and temperamental experiments easier, faster, and more consistent. Furthermore, a standardized, shareable method for preclinical CEST experiments with open-source data processing and reconstruction pipelines enables site-to-site comparison of CEST contrast measurements, potentially opening the door to multi-center trials assessing changes in myocardial composition and energetics in response to disease pathologies and treatments.

**ACKNOWLEDGEMENTS**

The instruments used in this work were supported and maintained by the Berkeley Preclinical Imaging Core. The authors would like to thank Jamie Walton for her assistance with adapting the cardiac segmentation code.
This work was supported by the National Institutes of Health (NIH) grant 5UH3EB028908-04.

**DATA AVAILABILITY STATEMENT**

The Python code used for image reconstruction and data processing is available at https://github.com/jweigandwhittier/cestsegCSUTE_scripts. Pulse sequence files are made available by request and through the Bruker Preclinical Imaging Community Forum: https://pci-community.com/. Sample datasets are available at https://doi.org/10.6084/m9.figshare.26112346.

**FIGURES AND TABLES**

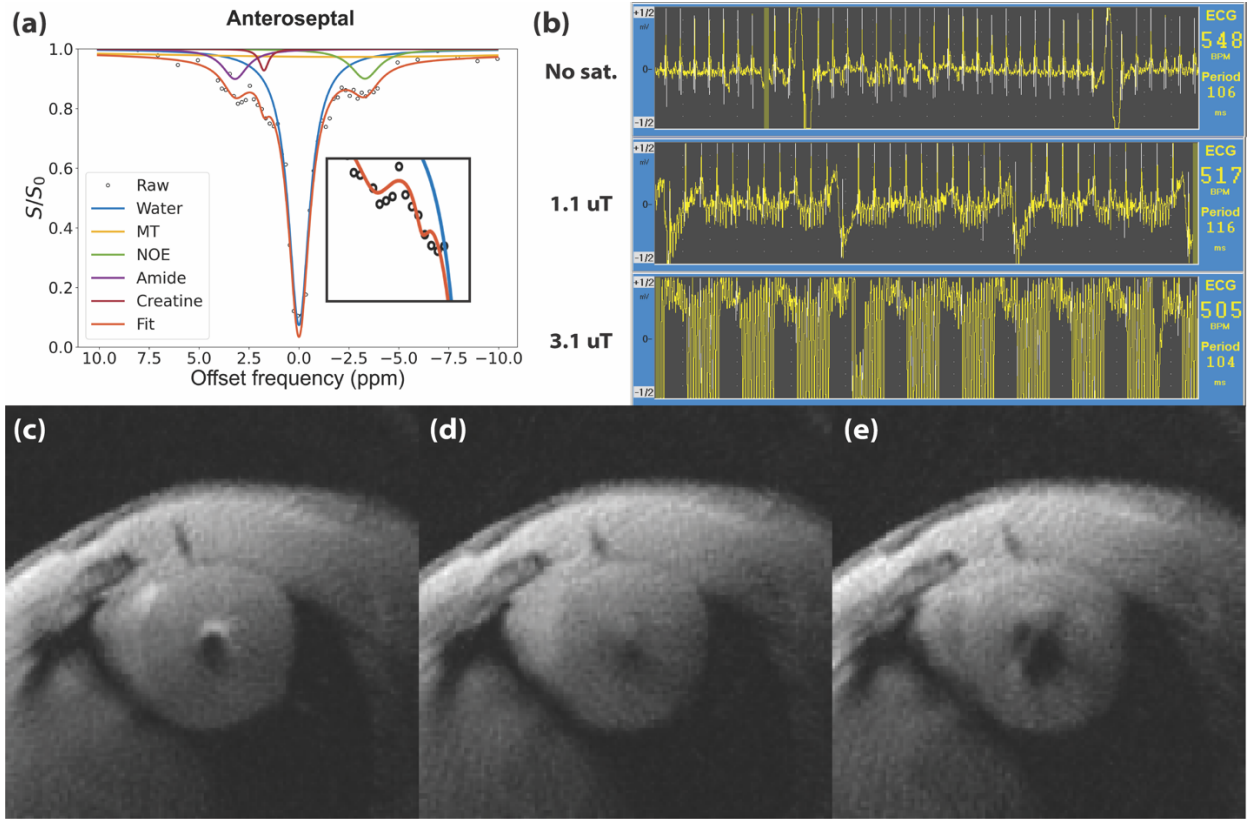

**Figure 1:** ECG and respiratory gating introduce inconsistencies in CEST acquisitions. Inconsistencies in Z-magnetization (a) are introduced by missed ECG triggers, which can be attributed to radiofrequency interference from high $B_1$ saturation pulses (b). As cardiac rates change throughout an acquisition, the captured cardiac phase and anatomy are altered within each offset and reference image, making image co-registration problematic. Representative reference images acquired at the start (c), middle (d), and end (e) of a single, gated acquisition capture distinct cardiac phases.

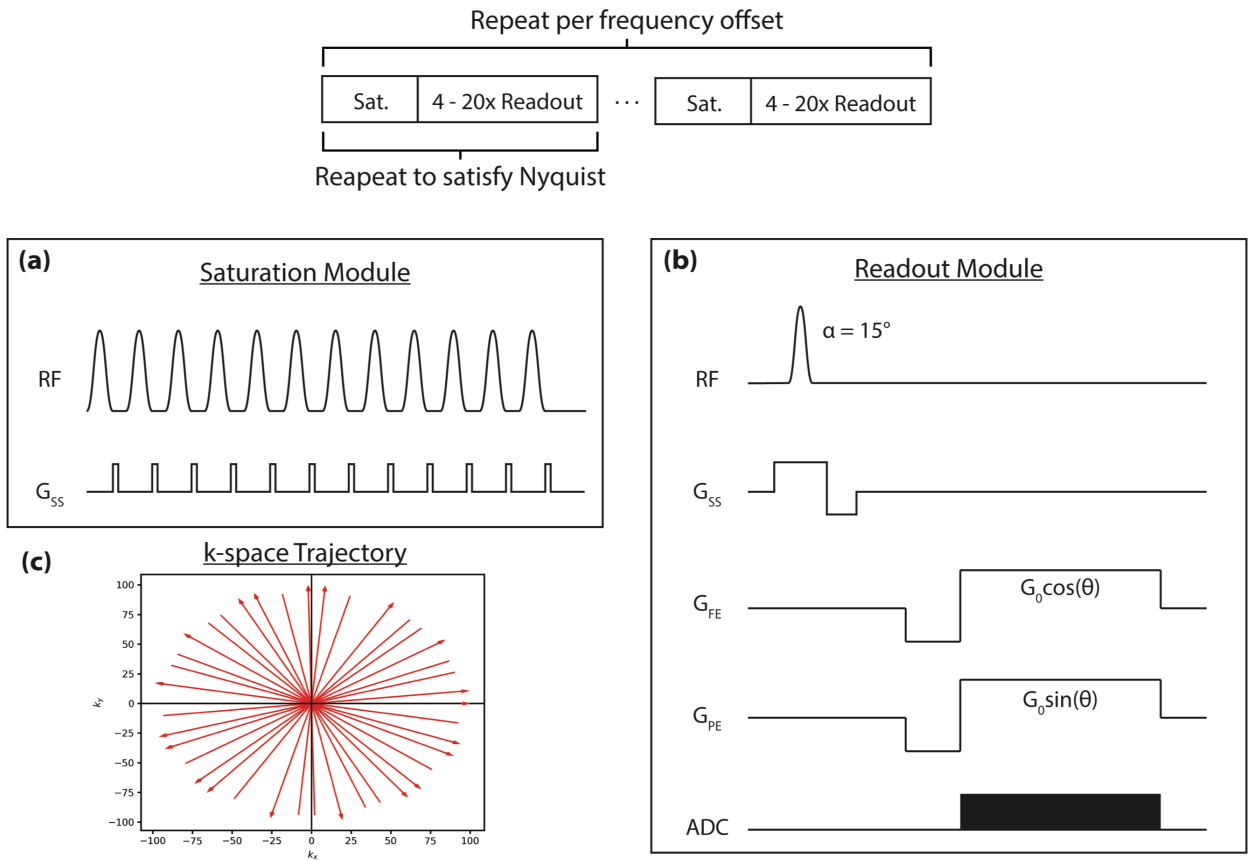

**Figure 2:** Ungated cardiac CEST sequence diagram.

A pulse sequence diagram for our modular, ungated preclinical cardiac CEST acquisition. Each sequence block consists of a short saturation period (a) followed by between 4 and 20 radial FLASH readouts (b, c). These saturation and readout blocks are repeated to satisfy up to 3.33x radial oversampling per image. The acquisition is then repeated at various frequency offsets, with a recovery time of 8 seconds, to generate the CEST Z-spectrum.

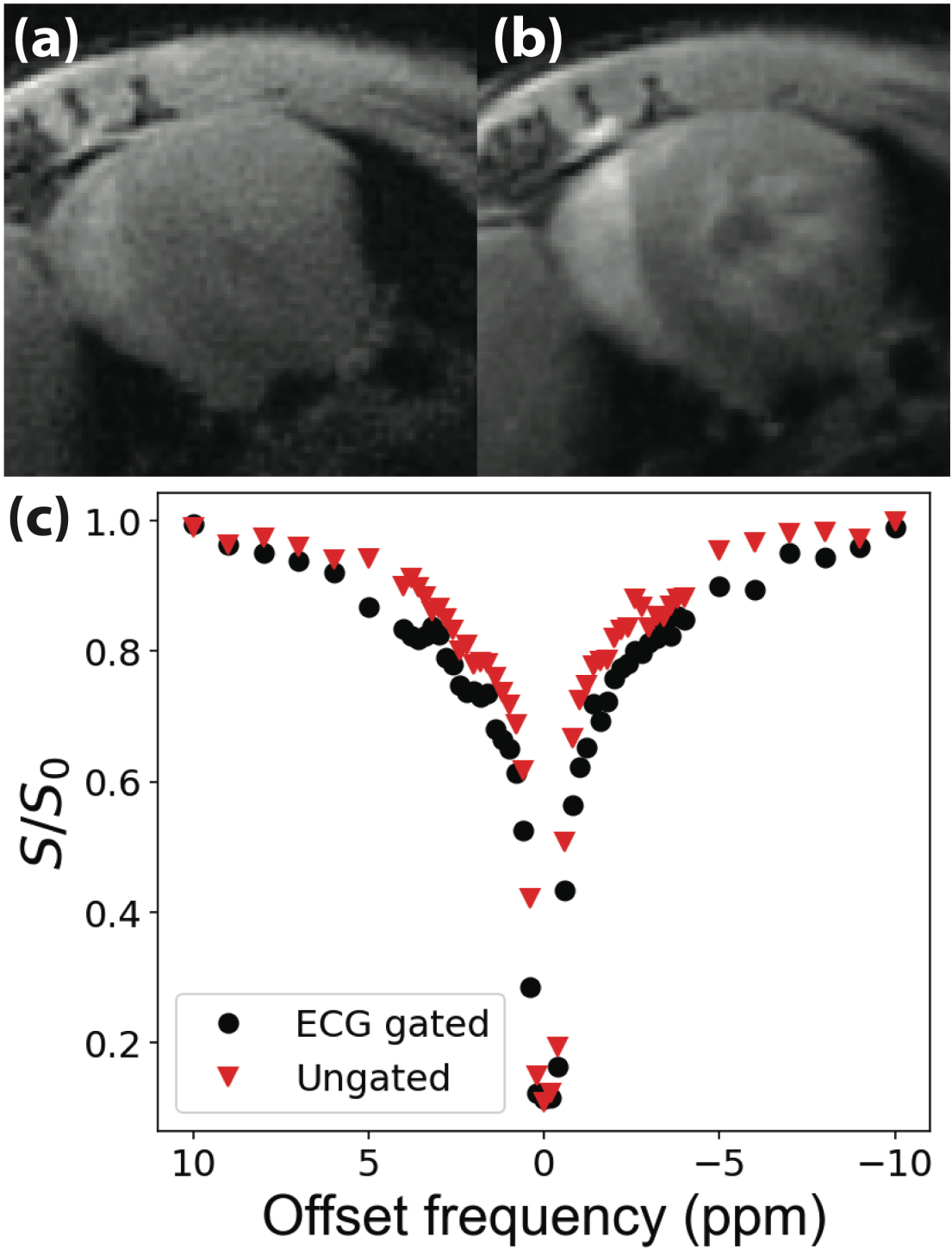

**Figure 3:** Representative reference images and Z-spectra from gated and ungated acquisitions. ECG gated (a) and ungated (b) reference images capture distinct cardiac phases relative to one another. In the ungated image, the myocardium and blood are clearly delineated. From the representative gated and ungated Z-spectra (c), the MT contribution is noticeably reduced in the ungated scenario. CEST contrasts are also slightly reduced.

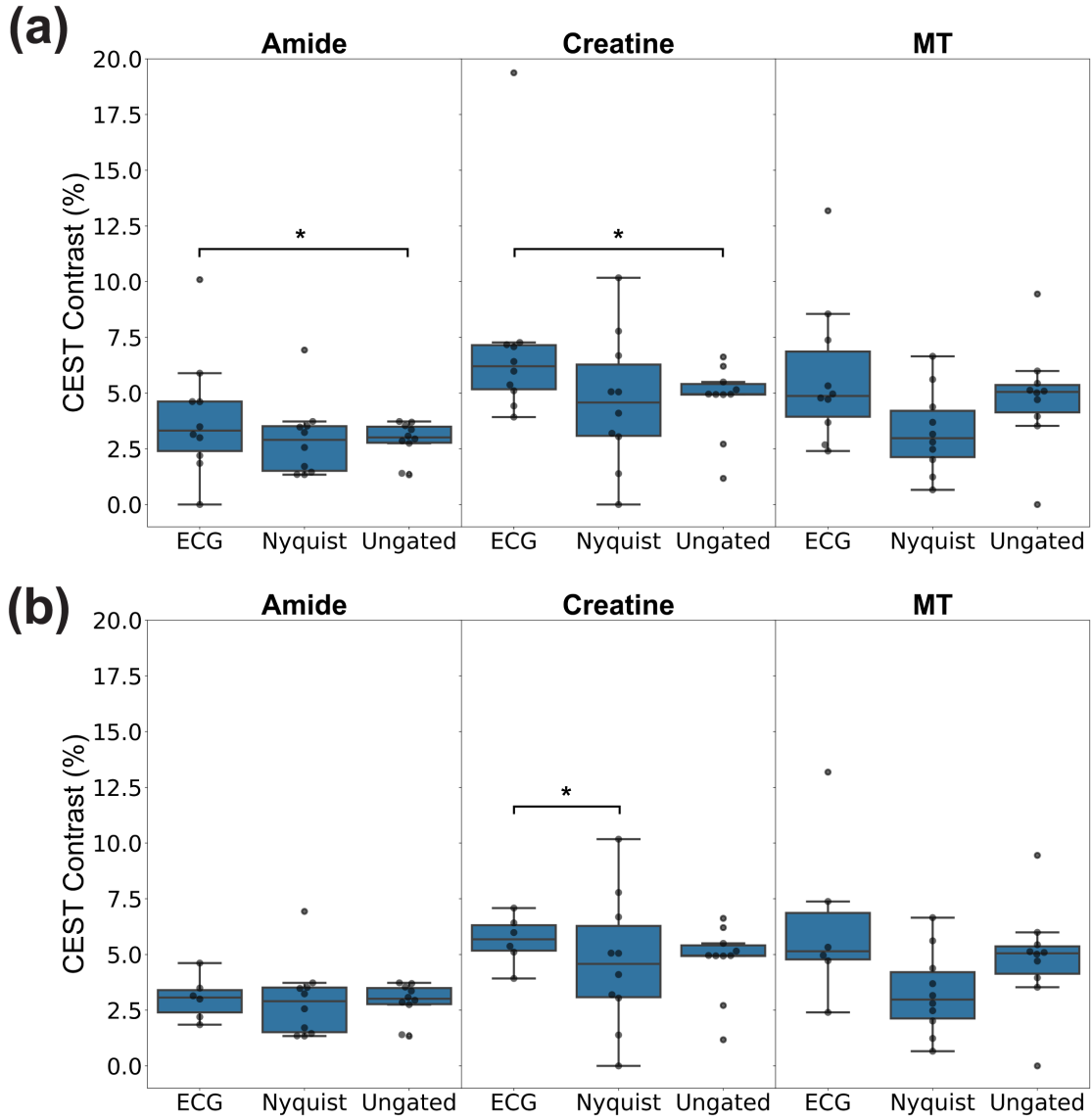

**Figure 4:** Combined results from CEST experiments comparing gated and ungated acquisitions. Box plots showing the distribution of measured CEST contrast with and without ECG gating and radial oversampling. With all acquisitions included (a), we observe broad interscan agreement the interscan agreement is significantly worse in the ECG-gated scenario. When acquisitions with poor ECG gating are excluded from the analysis (b), interscan variance decreased in the gated case and per-method differences are no longer significant. Radial oversampling improves interscan agreement in all cases.
* indicates a statistically significant difference in variance ($p < 0.05$).

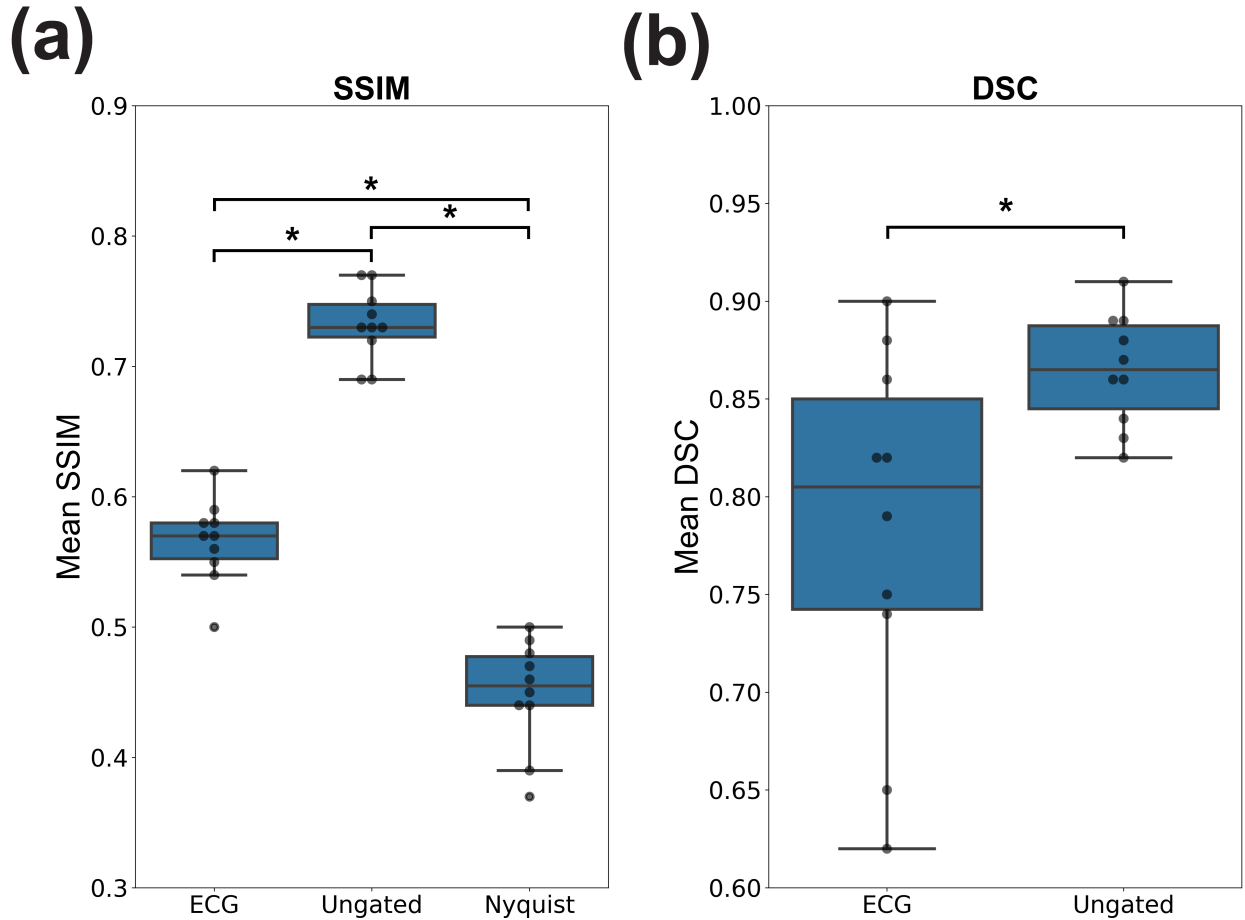

**Figure 5:** Quantitative analyses of image quality and differences in cardiac phase.

Box plots showing the distribution in mean SSIM between all CEST-weighted and reference images within each ECG gated and gated acquisition, with and without radial oversampling (a). A higher SSIM is correlated with better quality, more consistent images with superior SNR.

LV myocardial ROIs are compared using the Dice similarity coefficient throughout each acquisition. A higher mean DSC suggests more consistent cardiac phase throughout the acquisition (b).

* indicates a statistically significant difference between groups ($p < 0.05$).

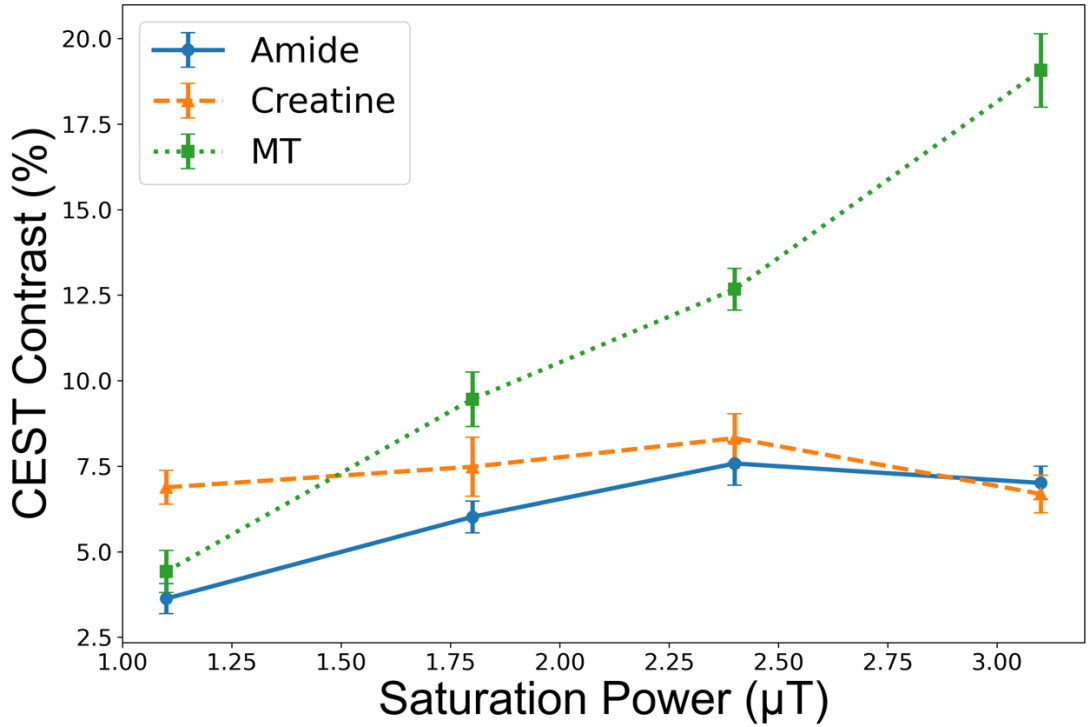

**Figure 6:** Comparative analysis of CEST contrasts at multiple saturation powers.

At 1.8μT, creatine and amide contrasts are both >5% on average, but distinct. At higher powers, the MT contribution is significant, and amide and creatine contrasts become difficult to separate due to peak broadening. Error bars indicate standard error.

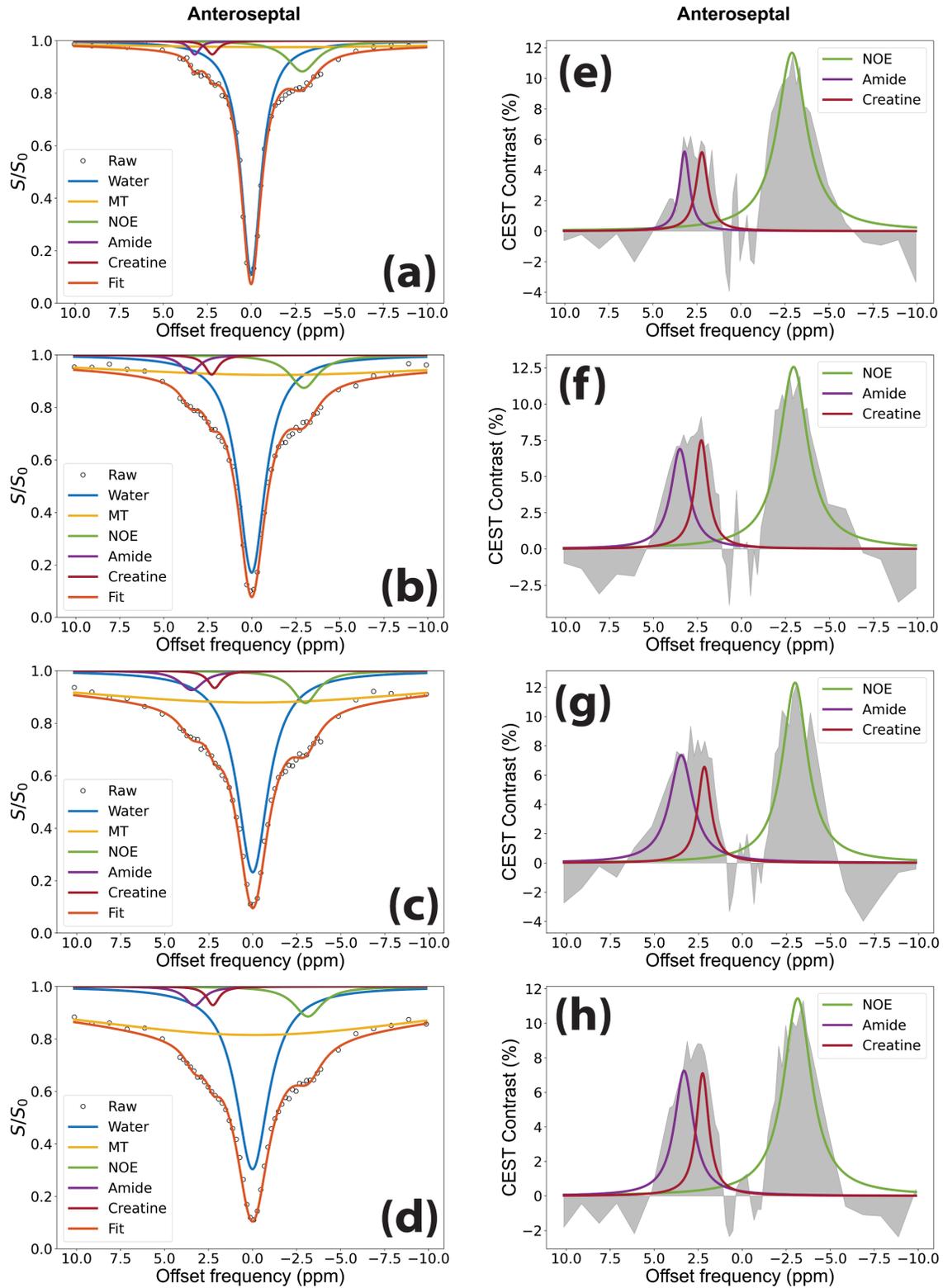

**Figure 7:** Representative Z-spectra, and Lorentzian difference plots at multiple saturation powers. Representative Z-spectra (a-d) and Lorentzian difference plots (e-h) from a single animal and imaging session with saturation powers of 1.1μT (a, e), 1.8μT (b, f), 2.4μT (c, g), and 3.1μT (d, h).

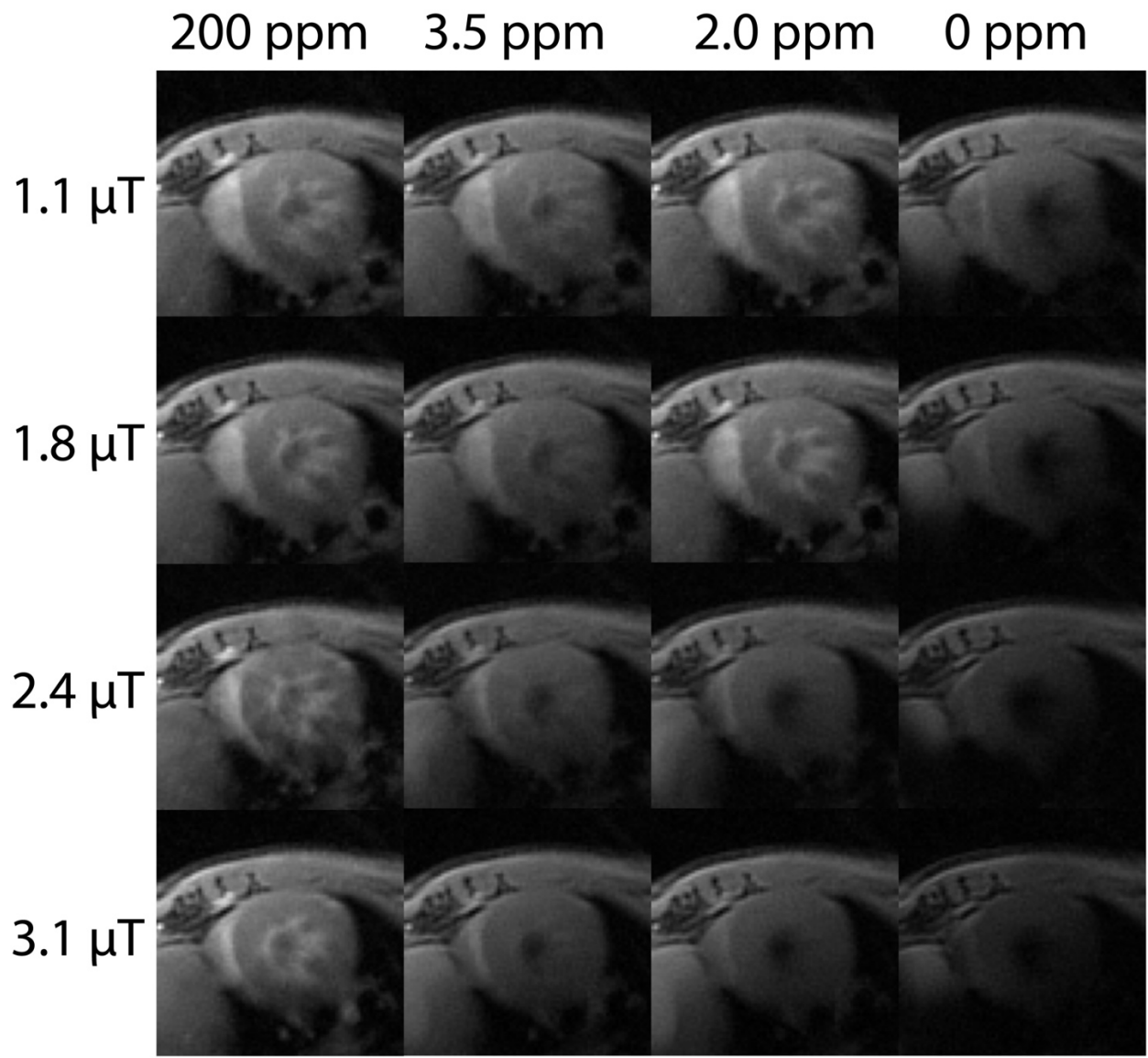

**Figure 8:** Representative images at relevant frequency offsets and saturation powers. Offset frequencies were calculated after $B_0$ correction in the anteroseptal segment. Images with water direct saturation at 0 ppm illustrate $B_1$ inhomogeneities, particularly as saturation powers are increased. Myocardial contrast and cardiac phase are consistent across all images.

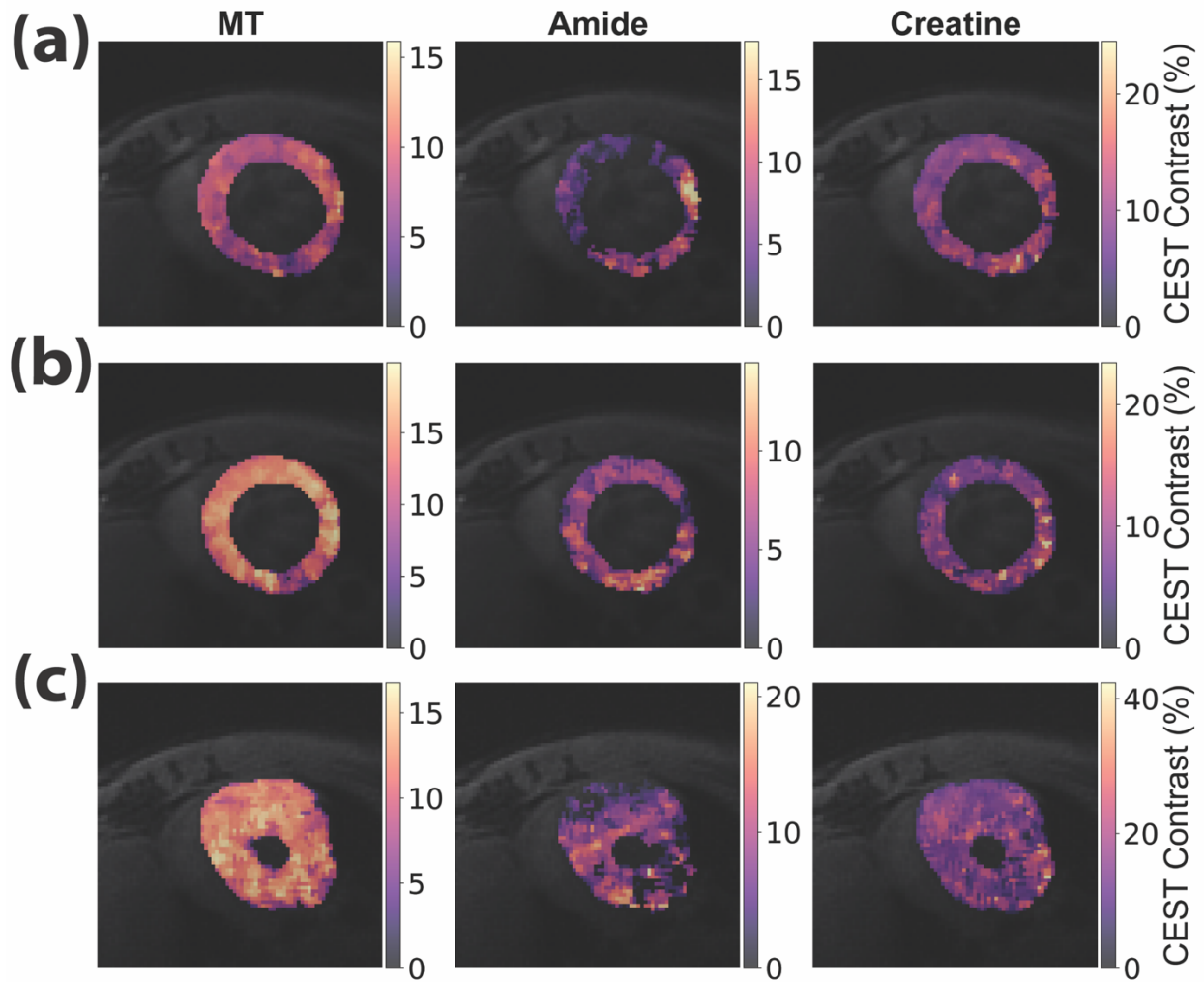

**Figure 9:** Pixelwise CEST contrast maps generated from images acquired using the ungated (a,b) and gated (c) sequences with saturation $B_1$ of 1.1µT (a,c) and 1.8µT (b).

Pixels exceeding the maximum number of iterations for Lorentzian fitting are zero-filled. After pixelwise fitting, contrast maps were processed using a median smoothing filter (kernel size = 3). CEST contrast homogeneity decreases significantly in inferior and lateral segments due to $B_0$ and $B_1$ inhomogeneity in the LV inferior wall.

|  | Water | MT | NOE | Amide | Creatine |
|---|---|---|---|---|---|
| Amplitude | [0.02, 1]; 0.8 | [0, 1]; 0.15 | [0, 0.25]; 0.05 | [0, 0.3]; 0.05 | [0, 0.5]; 0.05 |
| FWHM | [0.01, 10]; 0.2 | [30, 60]; 40 | [1, 5]; 1 | [0.2, 5]; 1.5 | [0.1, 5]; 0.5 |
| Center (ppm) | [-1e-6, 1e-6]; 0 | [-2.5, 0]; -1 | [-4.5, -1.5]; -2.75 | [3.2, 4.0]; 3.5 | [1.6, 2.6]; 2.0 |

**Table 1:** Parameters for Lorentzian-line-fit analysis by two-step least-squares fitting.

Least-squares fit parameters used for all Z-spectral fitting, presented as [lower bound, upper bound]; starting point. Water and MT are fit first; then amide, creatine, and NOE contrasts are fit to the Lorentzian difference: Z-spectral data with water and MT fits subtracted.

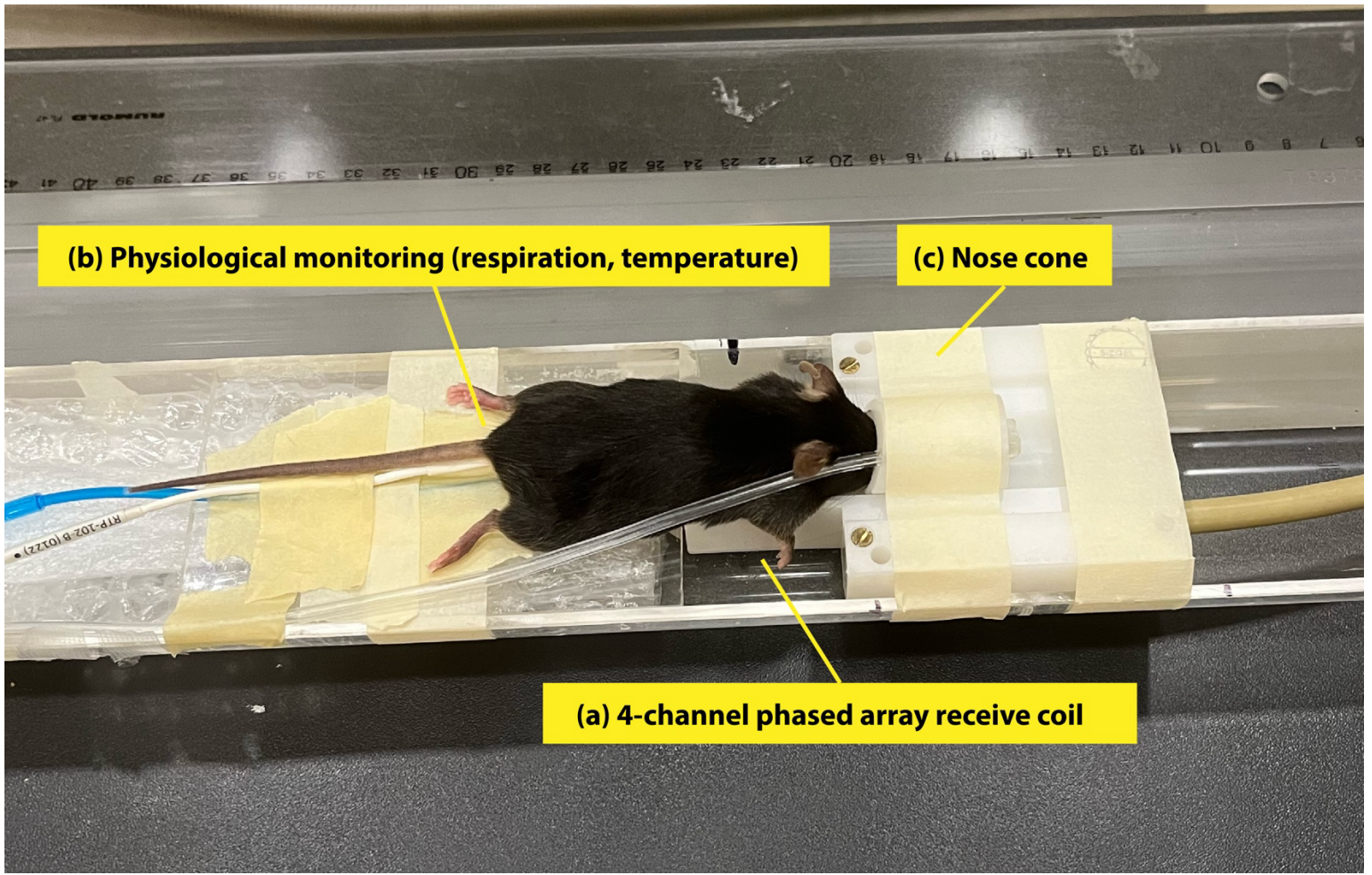

**Supporting Information Figure S1.** Experimental setup for ungated CEST acquisitions. Animals were oriented in the prone position with the thoracic region placed over the four-channel phased array receive coil (a). A pneumatic pillow and rectal thermistor temperature probe were used to monitor respiration and internal temperature (b). During gated acquisitions, ECG probes were also used (not pictured).

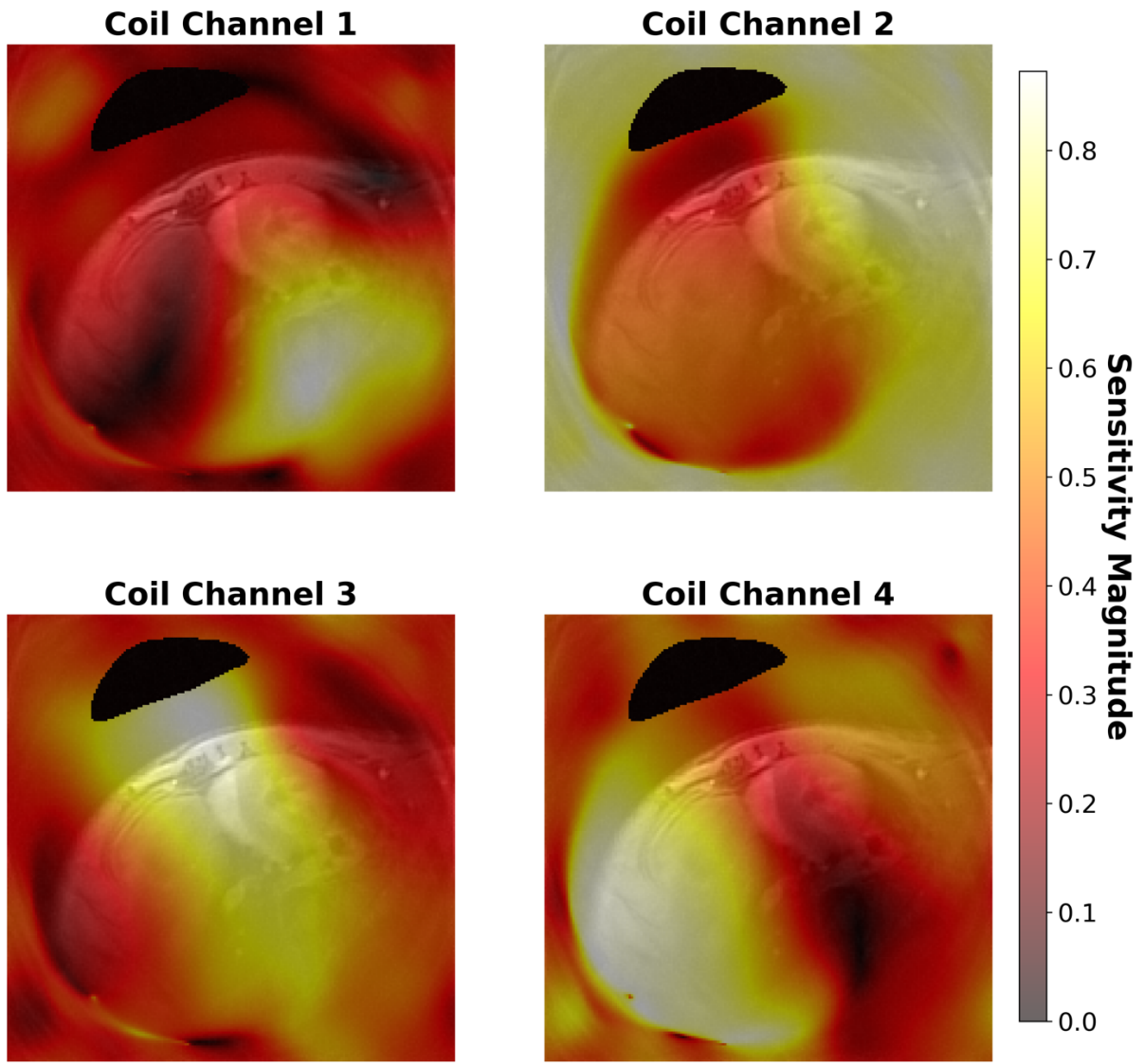

**Supporting Information Figure S2.** Per-coil sensitivity maps calculated with ESPIRiT using BART. The presence of apparent differences in direct saturation and CEST contrast quantification despite favorable receive coil sensitivity profiles suggests that susceptibility differences and $B_1$ inhomogeneities are the primary factors influencing observed Z-spectral inconsistencies.

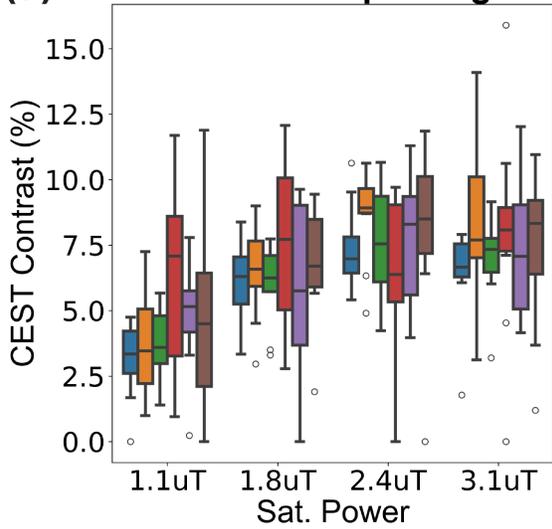
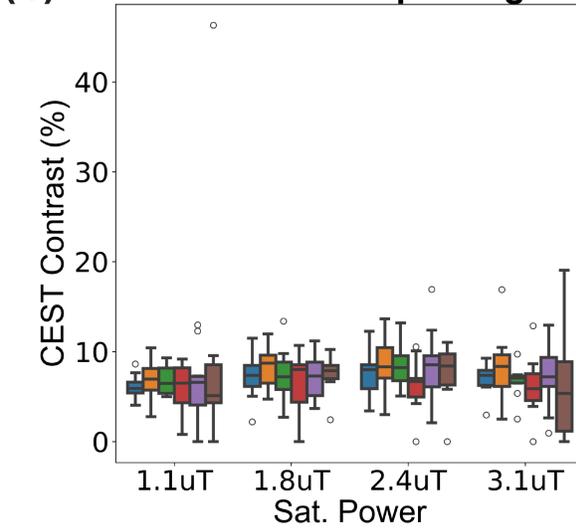
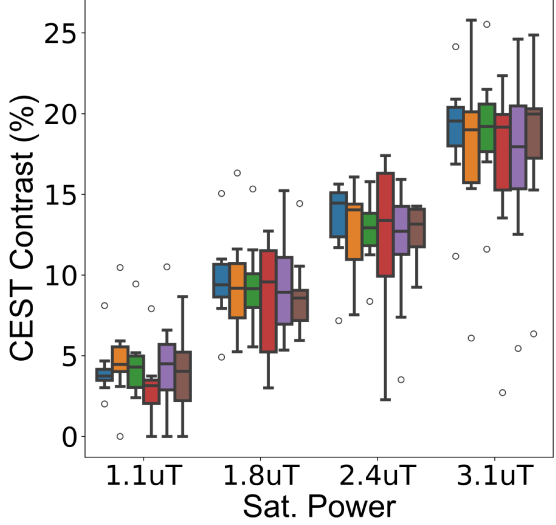
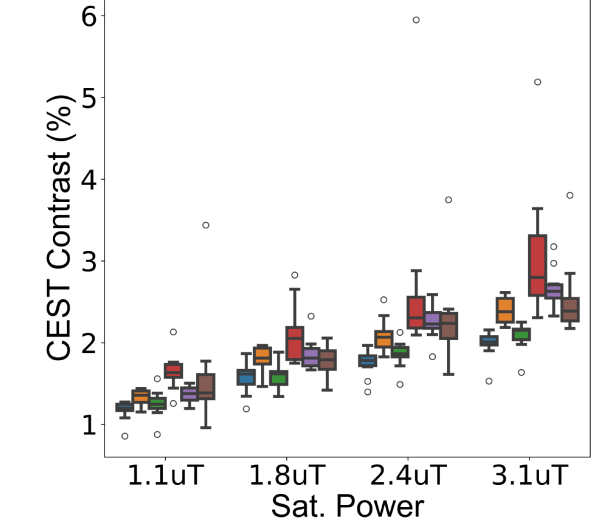
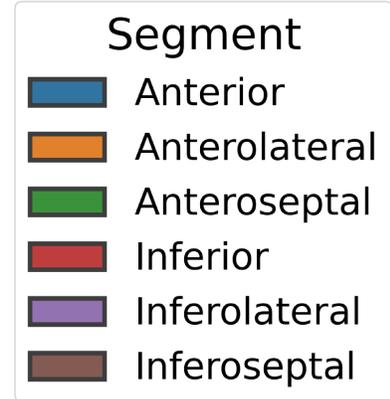

**Supporting Information Figure S3.** CEST contrasts and DS FWHM across cardiac segments at various saturation power. APT contrast (a), creatine contrast (b), MT contrast (c), and water direct saturation (DS) FWHM (d) per saturation $B_1$ and cardiac segment. Per-contrast variance and water direct saturation are increased in inferior and lateral segments due to $B_1$ inhomogeneity and changes in magnetic susceptibility.

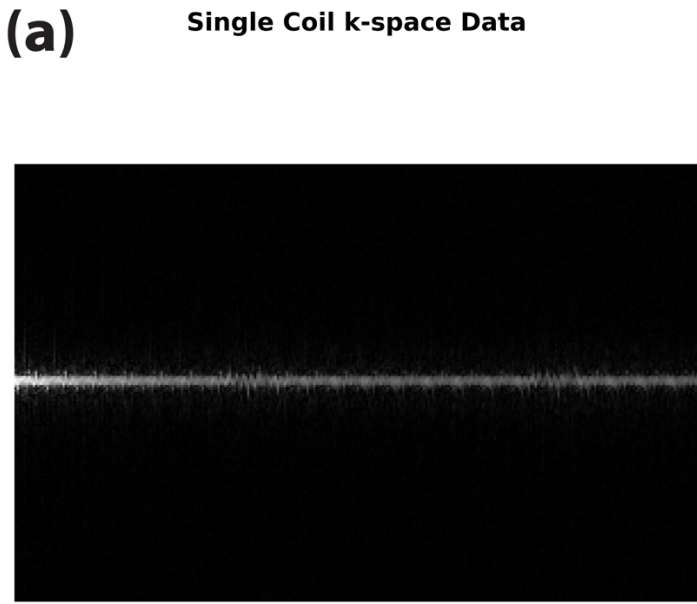
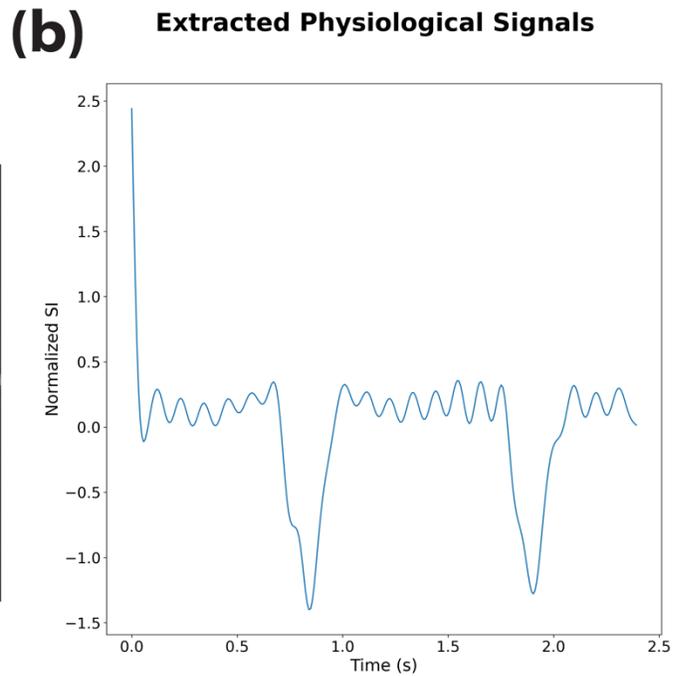
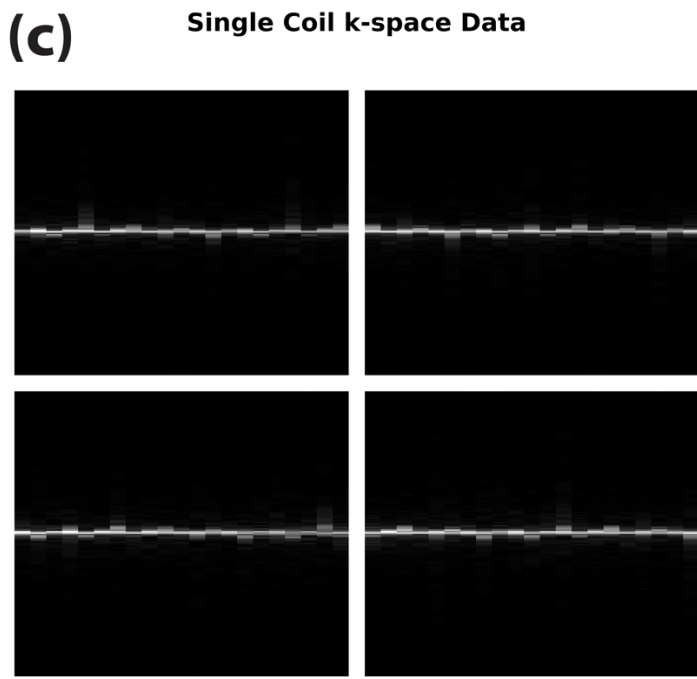
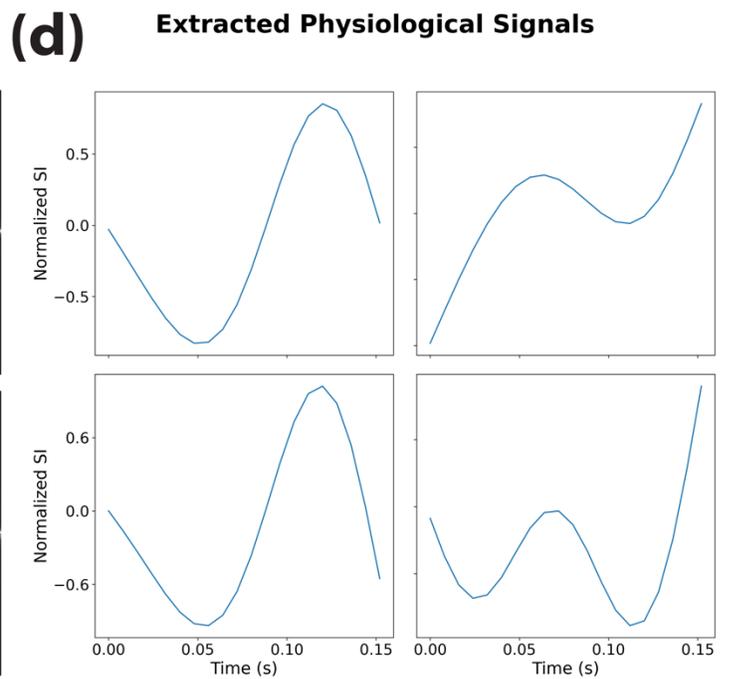

**Supporting Information Figure S4.** Cardiac and respiratory waveforms extracted from radial CEST data. Ungated radial acquisitions also allow for cardiac and respiratory waveforms to be extracted from raw *k*-space data. Center *k*-space points are isolated from each readout (a) and high-frequency components are filtered out. Without segmented saturation blocks, the entire cardiac and respiratory waveform can be extracted (b). When saturation blocks are employed (c), cardiac and respiratory waveforms can be reconstructed from each individual readout segment (d).

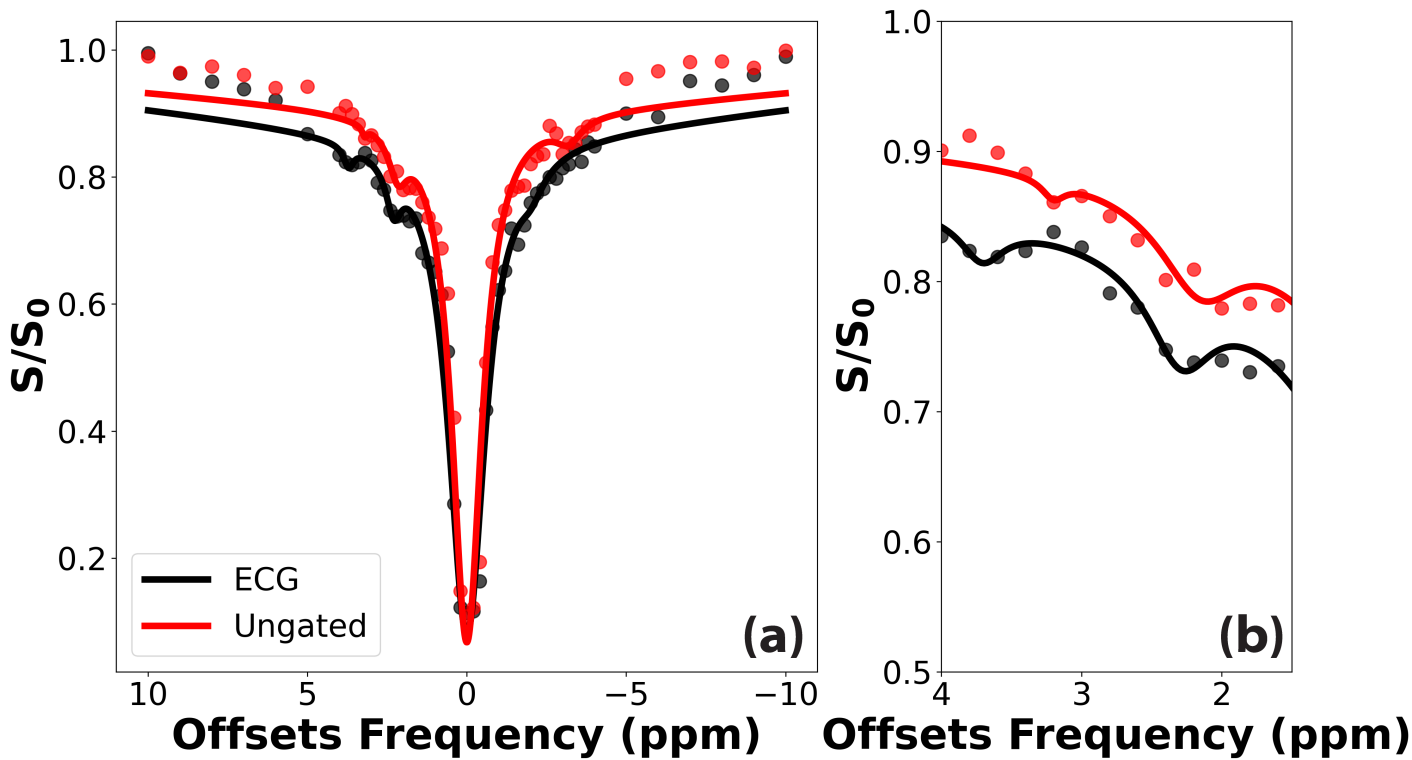

**Supporting Information Figure S5.** Representative reference images and Z-spectra from gated and ungated acquisitions (Figure 3c) with Lorentzian fits (a) and highlighted amide and creatine fitting regions (b).

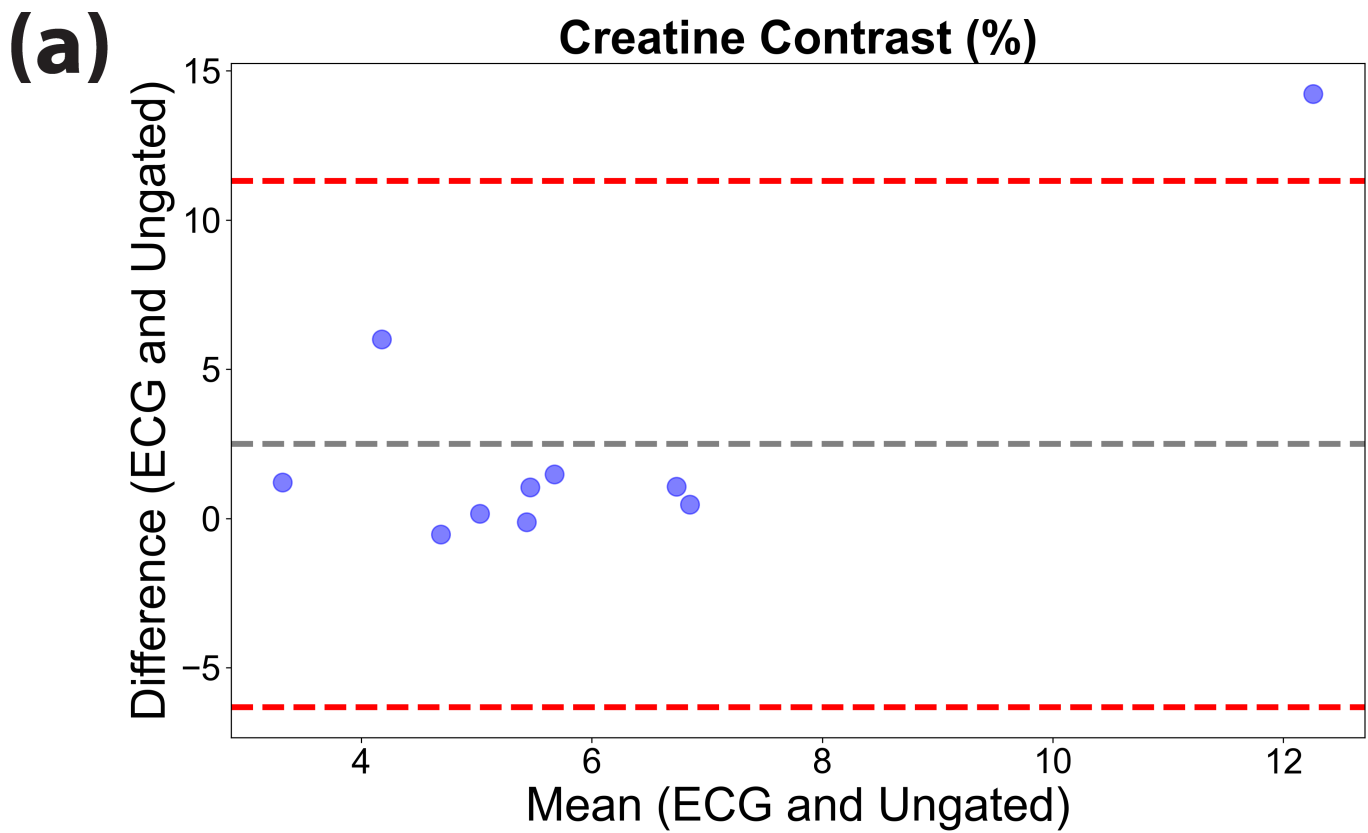

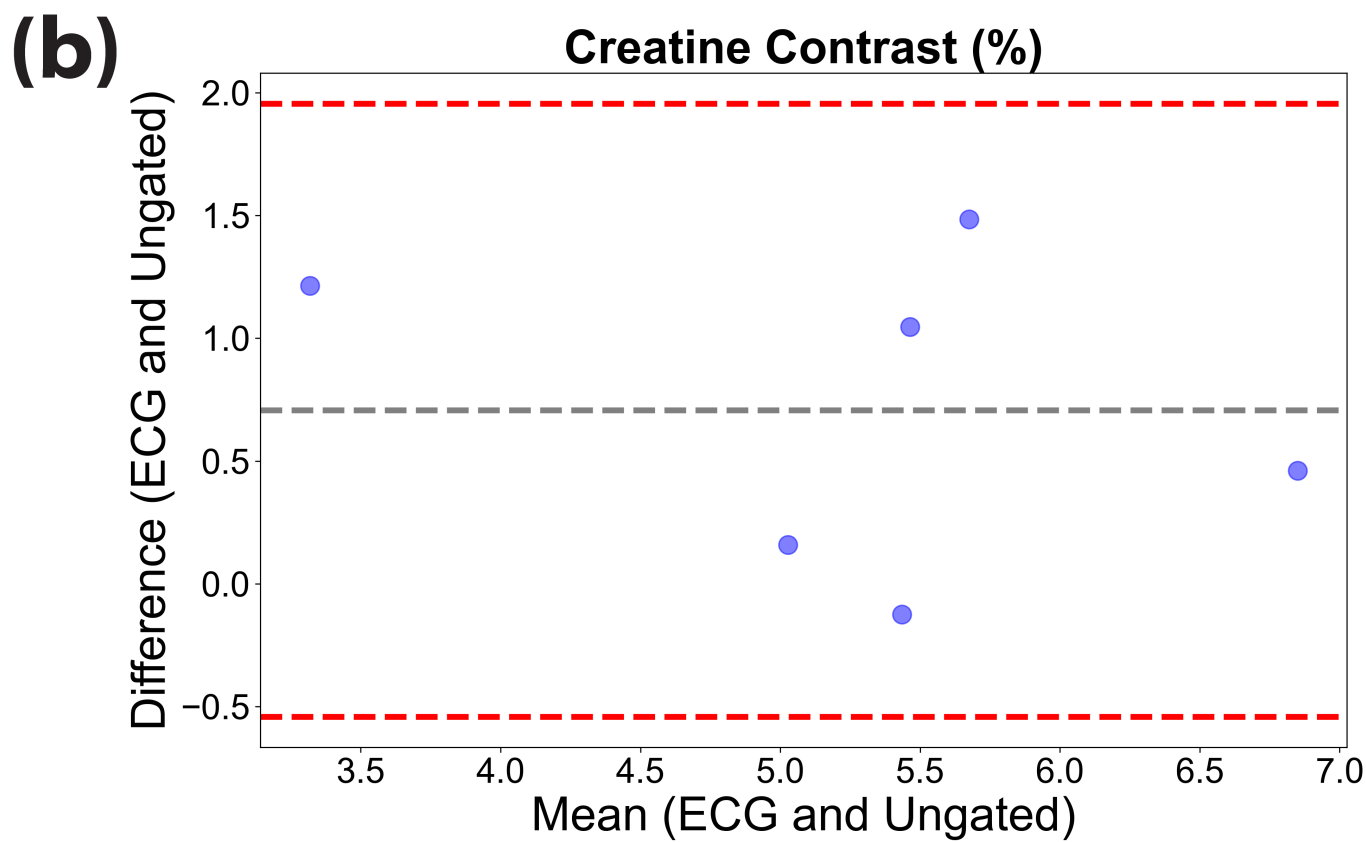

**Supporting Information Figure S6.** Bland-Altman plots characterizing the bias and mean difference in measured creatine contrast between ungated and ECG gated acquisitions with (a) and without (b) "poor gating" acquisitions included in the analysis.

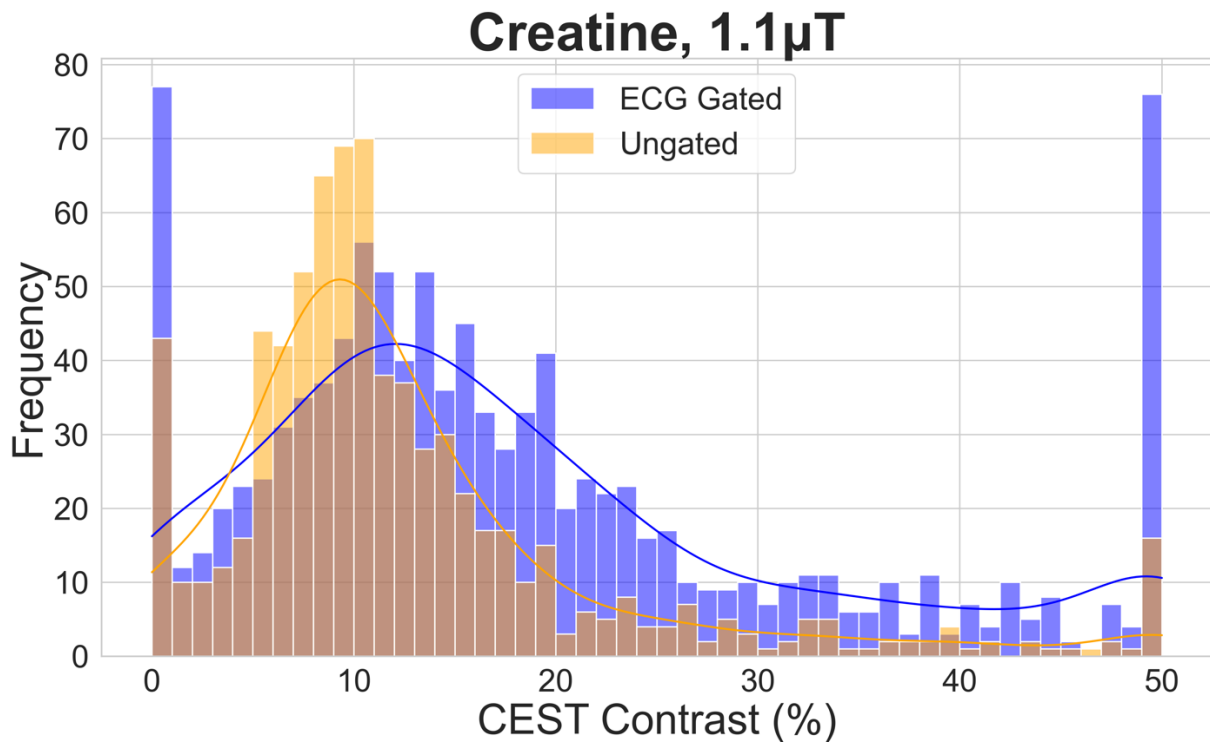

**Supporting Information Figure S7.** Histogram with kernel density estimation plot of fitted creatine values obtained from a single mouse using ECG gated and ungated sequences. CEST contrast maps generated from the ECG gated acquisition exhibit similar mean contrast values, with a greater percentage of erroneously fitted pixels clipped at maximum and minimum fit parameters. The standard deviation of the fits is also increased in the ungated scenario.